# ASYMPTOTICS OF THE CONTINUOUSDUAL HAHN POLYNOMIAL

A. D. Alhaidari[a], E. O. Ifidon[b] and T. J. Taiwo[b]

[a]*Saudi Center for theoretical physic, P.O. Box 32741, Jeddah 21438, Saudi Arabia*
[b]*Department of Mathematics, University of Benin, Benin city. Nigeria.*

**Abstract:** Using Darboux method, we obtain an asymptotic limit for the continuous dual Hahn polynomial. The basic concept is to construct a comparison function, which is the singular part of the generating function then expand the comparison function. Some simplification and manipulations result in the sought after asymptotics.

## 1. INTRODUCTION

The asymptotic limit $(n \to \infty)$ of an orthogonal polynomial $\{P_n\}$ is of great importance in quantum mechanics(nonrelativistic and relativistic) and in other related areas. For example, Geronimo [1] and Case and Geronimo [2, 3] explained the connection between scattering theory and asymptotics of orthogonal polynomials.These asymptotics enable one to obtain the scattering phase shift, bound states (finite and infinite) energies and resonances for a full description of the physical system.

Physical requirement, imply that the relevant class of orthogonal polynomials are those with the following asymptotic limit

$$P_n^\mu(\varepsilon) \approx A(\varepsilon)\cos\left[n^\xi \theta(\varepsilon) + \delta(\varepsilon)\right], (n \to \infty). \quad (0)$$

Where $A(\varepsilon)$ is the scattering amplitude at the energy $\varepsilon$ and $\xi$ is a real positive constant that dependson the particular energy polynomial. $\delta(\varepsilon)$ is the scattering phase shiftthat depends on the energy as well as the set of interaction parameters $\{\mu\}$ attached to the polynomial. On the other hand, bound states (if they exist) occur at the discrete real energies $\{\varepsilon_n\}$ at which $A(\varepsilon_n) = 0$ forcing the asymptotic wavefunction to vanish thus confining the system in configuration space. However, if $A(\varepsilon_n) = 0$ at complex energies $\{\varepsilon_n\}$ with negative imaginary part, then these are the resonance energies and the imaginary part forces the wavefunction to vanish with time due to the factor $e^{-i\varepsilon_n t/\hbar}$.

The major concern of this note is to derive the asymptotic limits of orthogonal polynomials (0). As an example, we obtain the asymptotics of the continuous dual Hahn polynomial using the Darboux method (chapter 8 of [4]).

## 2. ASYMPTOTIC LIMIT

To derive the large $n$ asymptotic formula of an orthogonal polynomial we apply the Darboux's method to its generating function as described in Sec. 9 of Chapter 8 in [4]. The concerned generating function is [5]

$$\sum_{n=0}^{\infty} \tilde{S}_n^{\mu}(y^2;a,b)t^n = (1-t)^{-\mu+iy} {}_2F_1\left({a+iy,b+iy \atop a+b}\bigg|t\right), \qquad (1)$$

where $t$ is in the complex plane and $\tilde{S}_n^{\mu}(y^2;a,b)$ is the continuous dual Hahn polynomial [5], which is defined here as

$$\tilde{S}_n^{\mu}(y^2;a,b) = \frac{(\mu+a)_n(\mu+b)_n}{n!(a+b)_n} {}_3F_2\left({-n,\mu+iy,\mu-iy \atop \mu+a,\mu+b}\bigg|1\right). \qquad (2)$$

For physical applications, we assume that all parameters are real and $y \geq 0$. The generating function (1) is holomorphic for $|t|<1$ and has a singularity at $t=1$. In our calculation, we need the contiguous relation (Eq. 7 of Ex. 21 in [6])

$$(a+b-c)\,{}_2F_1\left({a,b \atop c}\bigg|z\right) = a(1-z)\,{}_2F_1\left({a+1,b \atop c}\bigg|z\right) - (c-b)\,{}_2F_1\left({a,b-1 \atop c}\bigg|z\right), \qquad (3)$$

and the Euler transformation [6]

$${}_2F_1\left({a,b \atop c}\bigg|z\right) = (1-z)^{c-a-b}\,{}_2F_1\left({c-a,c-b \atop c}\bigg|z\right). \qquad (4)$$

Moreover, we will also apply the Gauss sum (Theorem 18 in §32 of [6])

$${}_2F_1\left({a,b \atop c}\bigg|1\right) = \frac{\Gamma(c)\Gamma(c-a-b)}{\Gamma(c-a)\Gamma(c-b)}, \qquad \text{Re}(c-a-b) > 0. \qquad (5)$$

The contiguous relation (3) makes the right-hand side of the generating function (1) as

$$\frac{(1-t)^{-\mu+iy}}{2iy}\left\{(a+iy)(1-t)\,{}_2F_1\left({a+1+iy,b+iy \atop a+b}\bigg|t\right) - (a-iy)\,{}_2F_1\left({a+iy,b-1+iy \atop a+b}\bigg|t\right)\right\}. \qquad (6)$$

We use the Euler transformation (4) to write the first term in the curled bracket as

$$(a+iy)(1-t)^{-2iy}\,{}_2F_1\left({b-1-iy,a-iy \atop a+b}\bigg|t\right) = (a+iy)(1-t)^{-2iy}\,{}_2F_1\left({a-iy,b-1-iy \atop a+b}\bigg|t\right), \qquad (7)$$

since ${}_2F_1\left({a,b \atop c}\big|z\right) = {}_2F_1\left({b,a \atop c}\big|z\right)$. This shows that the dominant term in a comparison function is

$$\frac{1}{2iy}\left\{(a+iy)(1-t)^{-\mu-iy}\,{}_2F_1\left({a-iy,b-1-iy \atop a+b}\bigg|t\right) - (a-iy)(1-t)^{-\mu+iy}\,{}_2F_1\left({a+iy,b-1+iy \atop a+b}\bigg|t\right)\right\} \qquad (8)$$

The Gauss sum (5) evaluates the first ${}_2F_1$ in (8) at $t=1$ as $\dfrac{\Gamma(a+b)\Gamma(1+2iy)}{\Gamma(b+iy)\Gamma(a+1+iy)}$ and the second ${}_2F_1$ as $\dfrac{\Gamma(a+b)\Gamma(1-2iy)}{\Gamma(b-iy)\Gamma(a+1-iy)}$. Therefore, Eq. (8) becomes

$$\frac{1}{2iy}\left\{(a+iy)(1-t)^{-\mu-iy}\frac{\Gamma(a+b)\Gamma(1+2iy)}{\Gamma(a+1+iy)\Gamma(b+iy)} - (a-iy)(1-t)^{-\mu+iy}\frac{\Gamma(a+b)\Gamma(1-2iy)}{\Gamma(a+1-iy)\Gamma(b-iy)}\right\} \qquad (8)'$$

Now, we use the gamma function identity $\Gamma(1+z) = z\Gamma(z)$ to write this as

$$(1-t)^{-\mu-iy}\frac{\Gamma(a+b)\Gamma(2iy)}{\Gamma(a+iy)\Gamma(b+iy)} + (1-t)^{-\mu+iy}\frac{\Gamma(a+b)\Gamma(-2iy)}{\Gamma(a-iy)\Gamma(b-iy)}. \qquad (8)''$$

Aside from $t$-independent factors, the comparison function near $t=1$ is $(1-t)^{-\mu-iy}$. The expansion of this term is

$$(1-t)^{-\mu-iy} = \sum_{n=0}^{\infty}\frac{(\mu+iy)_n}{\Gamma(n+1)}t^n. \qquad (9)$$

Therefore, the Darboux method states that

$$\tilde{S}_n^\mu(y^2;a,b) \approx \frac{(\mu+iy)_n}{\Gamma(n+1)} \frac{\Gamma(a+b)\Gamma(2iy)}{\Gamma(a+iy)\Gamma(b+iy)} + \text{complex conjugate}$$
$$\approx \frac{n^{\mu+iy-1}}{\Gamma(\mu+iy)} \frac{\Gamma(a+b)\Gamma(2iy)}{\Gamma(a+iy)\Gamma(b+iy)} + c.c. \qquad (10)$$

Where we have used $(z)_n = \frac{\Gamma(n+z)}{\Gamma(z)}$ and $\frac{\Gamma(n+a)}{\Gamma(n+b)} \approx n^{a-b}$. Writing a complex number $z$ as $z = |z|e^{i\arg(z)}$ and $z^* = |z|e^{-i\arg(z)}$, we can rewrite

$$\Gamma(2iy)/\Gamma(\mu+iy)\Gamma(a+iy)\Gamma(b+iy) =$$
$$|\Gamma(2iy)/\Gamma(\mu+iy)\Gamma(a+iy)\Gamma(b+iy)|e^{i\gamma} \qquad (11)$$

where $\gamma = \arg[\Gamma(2iy)/\Gamma(\mu+iy)\Gamma(a+iy)\Gamma(b+iy)]$. Therefore, we can write (10) as follows

$$\tilde{S}_n^\mu(y^2;a,b) \approx \frac{\Gamma(a+b)|\Gamma(2iy)|n^{\mu-1}}{|\Gamma(\mu+iy)\Gamma(a+iy)\Gamma(b+iy)|} \left(e^{iy\ln n} e^{i\gamma} + c.c.\right)$$
$$= \frac{2\Gamma(a+b)|\Gamma(2iy)|n^{\mu-1}}{|\Gamma(\mu+iy)\Gamma(a+iy)\Gamma(b+iy)|} \cos(y\ln n + \gamma) \qquad (12)$$

where we have also used $a^{ib} = e^{ib\ln a}$ and $e^{i\theta} + c.c. = e^{i\theta} + e^{-i\theta} = 2\cos\theta$.

Consequently, we conclude that the scattering amplitude $A(\varepsilon)$ is proportional to the inverse of the square root of the weight function of the continuous dual Hahn polynomial since the normalized version of the latter reads as follows [5]

$$\rho^\mu(y;a,b) = \frac{1}{2\pi} \frac{|\Gamma(\mu+iy)\Gamma(a+iy)\Gamma(b+iy)/\Gamma(2iy)|^2}{\Gamma(\mu+a)\Gamma(\mu+b)\Gamma(a+b)}. \qquad (13)$$

Additionally, since $\ln n \approx o(n^\xi)$ for any $\xi > 0$ then if we compare the general asymptotic formula (0) with Eq. (12) we conclude the scattering phase shift is $\delta(\varepsilon) = \gamma$. Finally, the bound states of the associated physical system is obtained from the conditions that make $A(\varepsilon) = 0$, which is satisfied if any one of the following conditions apply (depending on the values of the physical parameters): $\mu+iy = -n$, $a+iy = -n$, or $b+iy = -n$, where $n = 0, 1, 2, \ldots$ These conditions make the gamma function in the denominator of $A(\varepsilon)$ blow up since itsargument becomes non-positive integer. Thus, the corresponding energy spectrum formula is: $y^2 = -(n+\mu)^2$, $y^2 = -(n+a)^2$, or $y^2 = -(n+b)^2$, respectively.

## 3. CONCLUSION

In this note and with full details, we derived the asymptotic limit of the continuous dual Hahn polynomial using Darboux method. Acknowledging, the importance of the asymptotic limits in determining the details of a physical system in quantum mechanics, it is advised that efforts should also be made in getting the asymptotic limits of other orthogonal polynomials such as the Wilson polynomial, Meixner-Pollaczekpolynomial, and so on.